# Spatially indirect interfacial excitons in n-ZnO/p-GaN heterostructures

**Simran, Subhabrata Dhar***

Department of Physics, IIT Bombay, Mumbai-Maharashtra-400076, India.

*E-mail: dhar@phy.iitb.ac.in

**Abstract**

Electroluminescence properties of epitaxially grown n-ZnO/p-GaN pn-heterojunctions are investigated as functions of applied bias and temperature. The study reveals the existence of indirect interfacial excitons at sufficiently low temperatures. Electroluminescence feature associated with these excitons redshifts with increasing forward bias. It has been found that the binding energy of these entities can be controlled through applied forward bias and can even be made higher than that of the excitons in ZnO bulk (60 meV). However, formation of these excitons becomes unsustainable when either the applied bias or the temperature crosses a threshold. This has been explained in terms of leakage and thermal escape of electrons (holes) into GaN (ZnO) side. Calculations for the band diagram and the binding energy of these spatially indirect electron-hole coulomb-coupled entities are carried out. Theoretical results are found to explain the experimental findings quite well.

ZnO has direct band gap ($E_g$) of ~3.37 eV and large exciton binding energy of ~60meV, which make the semiconductor an attractive candidate for optoelectronis[1,2]. However, in spite of intense research for the past several years, controllable and reproducible p-type doping in ZnO is yet to be accomplished[3,4]. Heterojunctions of n-ZnO with other p-type layers can be an alternative to exploit all the merits of ZnO while bypassing the p-type doping challenge[1,5–9]. In this regard, GaN, which has a very similar lattice structure as ZnO with lattice mismatch <1.8% along c-direction[10] and where controllable p-type doping is possible[11–13], can arguably be a wonderful choice. Electroluminescence (EL) from n-ZnO/p-GaN heterostructures grown by different techniques has indeed been reported by several groups[14–18]. However, a detailed and systematic study to understand EL in this system is still lacking.

ZnO and GaN have type-II band alignment[19–21]. The two materials also have significantly different spontaneous polarizations along the *c*-direction, which results in a net positive charge accumulation at the interface, when a (0001) ZnO film is grown on (0001) GaN. This polarization field may cause quantum confinement of the conduction band electrons in the ZnO side of the junction. The combination of the band alignment and the polarization charge accumulation at the interface can lead to a unique type of bound indirect excitons formation at the interface, whose electron part is quantum confined along c-direction in the ZnO side, while the hole part stays at the GaN side of the junction. Like the interlayer excitons observed in homo-/hetero-bilayers of transition metal dichalcogenides (TMDs)[22–26], these spatially indirect excitons should also have planer character with a finite c-directional dipole moment. Here, the added advantage is the possibility to bring the plane of electron motion sufficiently close to that of the hole through applied forward bias. Certain important properties of these excitons, such as the transition energy, binding energy and g-factor can thus be modified through applied field. Due to the spatial separation between the electron and hole, these excitons are expected to show long lifetime, which could be useful for the development of exciton based logic circuits in the future[23]. Note that interlayer excitons in TMD heterostructures are found to show lifetime exceeding nanoseconds[24,25] and electrically tuneable resonance[26]. It should be further mentioned that Bose-Einstein condensation of these entities is theoretically possible as they possess finite perpendicular dipole moment [27,28]. A search for such excitons in ZnO/GaN heterojunction could thus be highly exciting.

Here, we investigate the behaviour of electroluminescence (EL) as a function of applied bias at different temperatures in pn-heterojunctions consisting of c-oriented n-ZnO layer grown epitaxially on c-GaN/sapphire template. In certain samples, an EL peak is found at low temperatures, which shows a redshift with increasing forward bias. Our study assigns this peak to spatially indirect interfacial excitons. Binding energy of these entities can be significantly increased by changing the applied bias and can even be made larger than that of ZnO bulk. Formation of such indirect electron-hole coupled systems is theoretically examined, which indeed supports the experimental outcome.

Here, we highlight two samples, where ~500 nm thick (0001) ZnO epitaxial films were deposited using pulsed laser deposition (PLD) technique on two different c-oriented p-type GaN/sapphire templates with room temperature hole concentrations of $n_h \sim 1 \times 10^{17}$ (sample D1) and $1 \times 10^{16} cm^{-3}$ (sample D2) in the GaN layer. ZnO layers are found to be unintentionally n-type doped with electron concentration of $n_n \sim 1 \times 10^{19}$ cm$^{-3}$ at room temperature. Details of the growth, structural and morphological properties of these films can be found in the supplementary information S1. Photoluminescence (PL) and electroluminescence (EL) measurements at different temperatures ranging from 10 to 300K were carried out in a closed cycle helium cryostat using a 550 mm focal length monochromator equipped with a UV-enhanced CCD detector. He–Cd laser of wavelength 325 nm was used as the excitation source for PL. Ni/Au (30nm/60nm) and Ti/Au (20nm/100nm) metal contacts were deposited on p-GaN and n-ZnO sides, respectively. Ni/Au contacts were rapid thermally annealed at 435ºC for sample D1 and at 400ºC for sample D2, in $N_2$ atmosphere for 5 minutes to get ohmic behaviour. Device structure is shown schematically in the inset of Fig.1(a). Current-voltage profiles recorded between the contacts on different sides were found to exhibit rectifying characteristics for both the samples down to the lowest temperature (10K) as discussed in the supplementary S2.

Fig.1(a) presents the EL spectra recorded for device D1 at 10K at different applied forward voltages. At sufficiently low voltages, the spectra are featured by two peaks at 3.21 and 3.12 eV. Intensity of both the peaks found by gaussian deconvolution of the combined feature (discussed in supplementary S3), increases with the applied bias. The lower energy peak does not show any shift with $V_b$, while the higher energy peak red-shifts as the bias voltage is increased as can be seen in Fig.1(b). Fig.1(c) shows the peak position variation of the 3.21 eV feature with $V_b$. Interestingly, the peak position shifts at a much slower rate for $V_b < 8V$, beyond which it decreases at a much faster pace. This low voltage regime can be termed as regime-(I). As $V_b$ is increased further, a doublet structure starts to appear at 3.10 and 3.15 eV as can be observed in Fig.1(a). Beyond a threshold voltage of around 10V, the intensity of this doublet feature rapidly increases and fully overwhelms the 3.21 and 3.12 eV features, which marks the beginning of regime-(II). In this regime, a weaker feature also appears around 2.4 eV. For $V_b > 12.5$ V, intensity of the ~3eV band decreases sharply, while its shape changes significantly with increasing bias. In this regime-(III), 3.1 eV doublet structure disappears. Instead, a peak appears at ~3.27 eV with a low energy hump at 2.9 eV. Interestingly, 2.4 eV feature continues to grow without showing any peak-shift. Fig.1(d) shows the 10K EL spectra recorded at different applied forward voltages for the sample D2. Evidently, all spectra are featured by three peaks. Two higher energy humps can be seen at 3.23 and 3.14 eV, while the most dominant peak appears at ~2.78 eV. The 2.78 eV feature shows a clear blue shift with the increasing bias, which may suggest its donor-acceptor-pair (DAP) origin. Unlike in device D1, none of the 3.23 and 3.14 eV peaks show any shift with $V_b$ in this sample. Note that comparisons between 10K EL spectra and PL spectra recorded separately on ZnO film and the GaN template for both the samples does not find any match of the positions of the EL peaks appearing above 3 eV with any of the near band edge features of either GaN or ZnO. This strongly suggests that these EL transitions must be originating from the interfacial region (see supplementary S3).

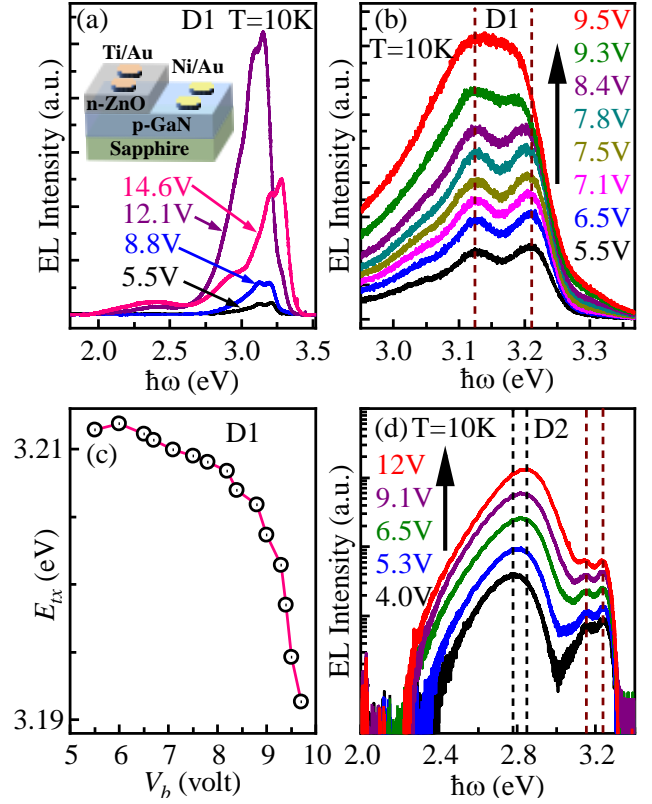

**Fig.1:** (a) EL spectra for the device D1 recorded at 10K under different applied forward biases. Inset: schematic view of the devices used in this study. (b) Highlights the bias dependent spectral change in regime-(I) for the same device. (c) Position of the 3.21 eV feature as a function of the applied bias $V_b$ in regime-(I). (d) EL spectra recorded at 10K under different applied biases for the device D2.

Note that our ZnO layers are deposited without any oxygen treatment of Ga-polar (0001) GaN surface before the PLD growth. Such a pre-treatment has been found to be necessary to grow O-polar (000$\bar{1}$)ZnO film on c-GaN substrates[29,30]. ZnO epitaxial layers are thus expected to be Zn-polar. As mentioned earlier that both GaN and ZnO have spontaneous polarization $P_{sp} = -0.029$ and $-0.05$ C/m$^2$, respectively[31,32] along [0001] direction. As a result, accumulation of a net positive charge of 0.021 C/m$^2$ is expected at the interface as shown schematically in Fig.2(a). Since XRD study does not evidence biaxial strain in these ZnO films (see supplementary S1), piezoelectric polarization at the interface can be excluded in our samples. Here, the energy band diagrams across the junction between Zn-polar n-ZnO grown on Ga-polar p-GaN are obtained by solving poisson equation with appropriate boundary and interface conditions considering the effects of spontaneous polarization and band off-set at the interface. More details of the calculation and the parameters used for the same are provided in supplementary S4.

Fig.2(a) presents the band diagram calculated at 10K for bias voltages $V_j = 0$ and 2V at the junction in case of the ZnO/GaN system with the acceptor concentration of $N_{aG} = 2 \times 10^{18}$ cm$^{-3}$ in the GaN side and donor concentration of $N_{dZ} = 1 \times 10^{19}$ cm$^{-3}$ in the ZnO side. Formation of the polarization induced triangular potential well adjacent to the junction for the electrons in the ZnO side and a depletion region in the GaN side is evident in both the cases. The potential well can quantum mechanically confine electrons as the solution of the Schrodinger equation returns eigen states below the fermi level even up to $V_j \sim 3V$. In Fig. 2(b), the conduction band diagrams in a close vicinity of the junction are compared for $V_j = 0$ and 2V. Note that for better comparison of the shape of the potential wells in the two cases, the band energy in the ZnO side at the junction has been treated as the reference in the figure. Ground state wavefunctions and energies (represented by the position of the base of the respective wavefunctions) are also shown in the figure for the two bias voltages.

Beyond a certain forward bias voltage, hole concentration near to the interface starts to build up. If the temperature is sufficiently low, the valence band off-set can restrict the holes to jump over to the ZnO side. In this scenario, ground state electrons confined in the triangular potential well in the ZnO side can be coupled with the holes to form the indirect excitons (IDX), in which electron part stays in a parallel plane that is $d_x$ distance away from the interface and the hole part is located in the GaN side adjacent to the junction. This is shown schematically in the inset of Fig.2(a). One can consider $d_x = <x>$, the expectation value of position of the ground state electron in the triangular potential well. As the forward bias is increased, the well narrows leading to an upshift of the ground state energy as can be seen in Fig.2(b). This results in the reduction of $d_x$, which is also evident from the figure. Binding energy $E_{bx}$ of such excitons can thus be controlled through applied forward bias. $E_{bx}$ can be obtained by solving the Schrodinger equation in cylindrical coordinate considering electron-hole interaction potential as $V(r) = -\alpha e^2/4\pi\epsilon_o \kappa_{av}\sqrt{\rho^2 + d_x^2}$, where $\kappa_{av} = (1/2)(\kappa_{GaN} + \kappa_{ZnO})$ the average of the dielectric constants of GaN and ZnO, $\epsilon_o$ the vacuum permittivity and $\alpha$ is a constant introduced to take into account the overestimation of dielectric screening effect in GaN and ZnO[33].

More details of the calculation can be found in supplementary S5. Transition energy of such excitons can be expressed as $E_{tx} = E_{c1}^{ZnO}(0) - E_v^{GaN}(0) - E_{bx}$, where $E_{c1}^{ZnO}(0)$ and $E_v^{GaN}(0)$ represent the ground state energy of the potential well in the ZnO side and the energy at valence band maximum in the GaN just at the junction. In Fig. 2(c), calculated $E_{bx}$ and $E_{tx}$ are plotted as functions of the applied bias at the junction $V_j$. As expected, $E_{bx}$ increases with $V_j$. Note that $E_{tx}$ versus $V_j$ profile shows quite a similar behaviour as that is experimentally observed for the sample D1 in Fig. 1(c). $\alpha$ is taken to be 2 in these calculations[36]. Note that the applied bias across the two contact pads $V_b$, which is plotted along x-axis in Fig. 1(c), should not be the same as the junction bias $V_j$ plotted in Fig. 2(c). In reality, $V_j$ is expected to be less than $V_b$. Nevertheless, a similarity in the shape of the two plots strongly support the assignment of 3.21 eV transition observed in sample D1 to IDX.

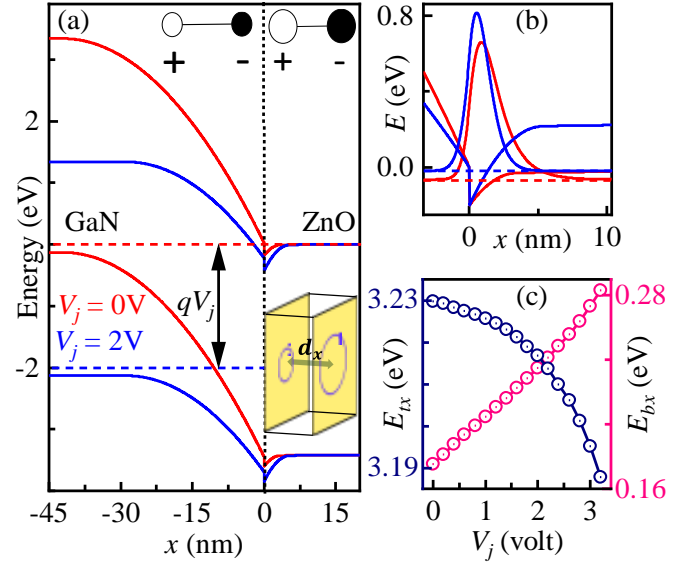

**Fig.2:** (a) Band diagrams calculated for two different applied biases. Spontaneous polarization directions in GaN and ZnO are schematically shown at the junction. Inset: schematic representation of the indirect exciton IDX. (b) Conduction band profile at the junction in an expanded scale. (c) Transition energy and binding energy of IDX as a function of applied bias at the junction.

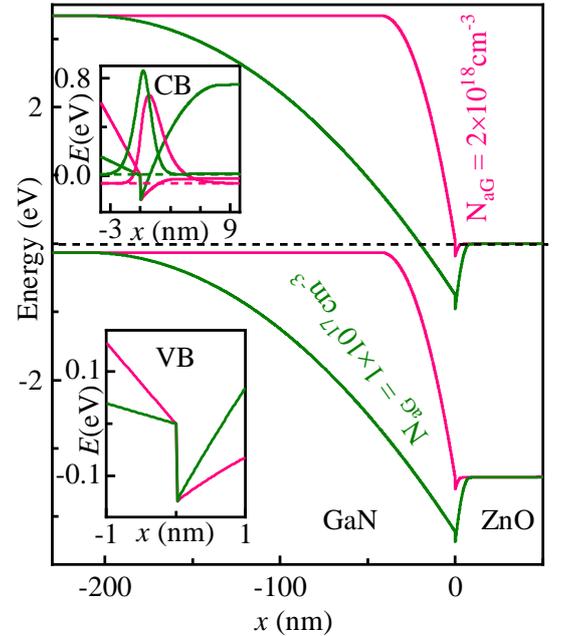

**Fig.3:** Band diagrams for different acceptor concentrations $N_{aG}$ in GaN side. Insets compare the conduction and valence band profiles at the junction.

In Fig.3, band diagrams calculated at 10K for $V_j = 0$ in case of the acceptor density $N_{aG} = 1 \times 10^{17}$ and $2 \times 10^{18}$ cm$^{-3}$ in the GaN side, are compared. Clearly, even at zero bias condition, the potential well (barrier) for the electrons (holes) is significantly narrower in case of lower $N_{aG}$ (see insets of the figure), which reduces the chance of IDX formation. This can explain why no EL peak showing red-shift with applied bias could be detected in sample D2, where $N_{aG}$ is indeed of the order of $10^{17}$ cm$^{-3}$.

We believe that the 3.23 eV EL peak in sample D2 [see Fig. 1(d)] is arising from transition between the shallow donor states in ZnO to valence band edge in GaN ($E_{dZvG}$). While, 3.14 eV EL peak can be attributed to conduction band (or shallow donors) to Mg acceptor states in GaN ($E_{cGaG}$) side of the junction[34]. The most dominant EL feature appearing at 2.78 eV in this sample can be assigned to blue luminescence (BL) band often reported in low temperature PL of Mg doped GaN (see supplementary S3). Assignment of different transitions are schematically shown in Fig. 4(a). In sample D1 with $N_{aG} \sim 2 \times 10^{18}$ cm$^{-3}$, IDX formation is theoretically sustainable and hence 3.21 eV IDX-feature can be observed at low temperatures. In this sample, IDX transition must be dominating over $E_{dZvG}$ transition in regime-(I). But, $E_{cGaG}$ peak is visible [Fig. 1(a)]. As the forward bias increases, the triangular potential well (barrier) in the conduction (valence) band of the ZnO side narrows, which increases the binding energy of the IDX, but at the same time reduces their chance of formation by enhancing the leakage of the electrons (holes) into GaN (ZnO). Above a certain threshold bias, IDX formation becomes unsustainable and various other transitions start to dominate mainly in GaN side (as holes are less mobile than electrons). We believe that the rapid increase of 3.10 eV peak beyond 10V of applied bias is the point when electrons and holes escapes into GaN and ZnO sides in sample D1 [see Fig. 1(a)]. Since sample D2 is always in regime-(II), sudden rise of any EL peak as a function of the bias could not be seen there. Increase of the applied bias in regime-(II) results in more electrons leaking into GaN side that leads to the enhancement of BL and YL transitions as observed in both the samples [see Fig. 1(a) and (d)]. At sufficiently high bias [regime-(III)], an EL peak emerges at 3.27 eV in sample D1. This can be assigned to $E_{dZvG}$. In this regime, EL yield reduces with increasing bias, which can be attributed to Auger recombination process[35].

Fig. 4(b) shows the evaluation of the EL spectra recorded at a bias of 8.8V with the increase of temperature in sample D1. The two peaks at 3.12 ($E_{cGaG}$) and 3.21 eV (IDX) are dominant only below ~60K. Above 60K, 3.10 eV doublet feature rapidly increases marking the onset of thermal escape of the electrons from the potential well. This leads to the sudden influx of electrons in the GaN side. However, for $T > 100$K, overall EL yield starts to drop, which can be attributed to nonradiative recombination processes. At higher temperatures, transitions can be observed even in the ZnO side indicating thermal escape of the holes over the junction barrier into the ZnO side (see supplementary S3). Fig.4(c) presents the EL spectra recorded below 60K. Inset shows the variation of the integrated intensity of the $E_{cGaG}$ and IDX features as functions of temperature. It is worth noticing that both the peaks initially increase as the hole population at the junction rises with temperature. At higher temperatures, the intensity of $E_{cGaG}$ peak reaches a plateau, while that of the IDX decreases with increasing temperature. This can be attributed to the dissociation of IDX due to the thermal escape of the holes over the junction barrier. In Fig.4(d), IDX intensities ($I_{IDX}$) recorded for different bias voltages are plotted with the inverse of temperature ($1/T$). Binding energy ($E_{bx}$) of the IDX, which has been obtained by fitting the data using equation $I_{IDX} = I_o/[1 + C \exp(-E_{bx}/k_BT)]$ with $I_o$ and $C$ are temperature independent constants, are plotted versus $V_b$ in the inset. Evidently, $E_{bx}$ increases with the applied forward bias and can even cross the excitonic binding energy value of ~60 meV for bulk ZnO[37]. Note that our theoretical calculations also predict large binding energy values, which increase with bias, for these excitons [see Fig. 2(c)][37].

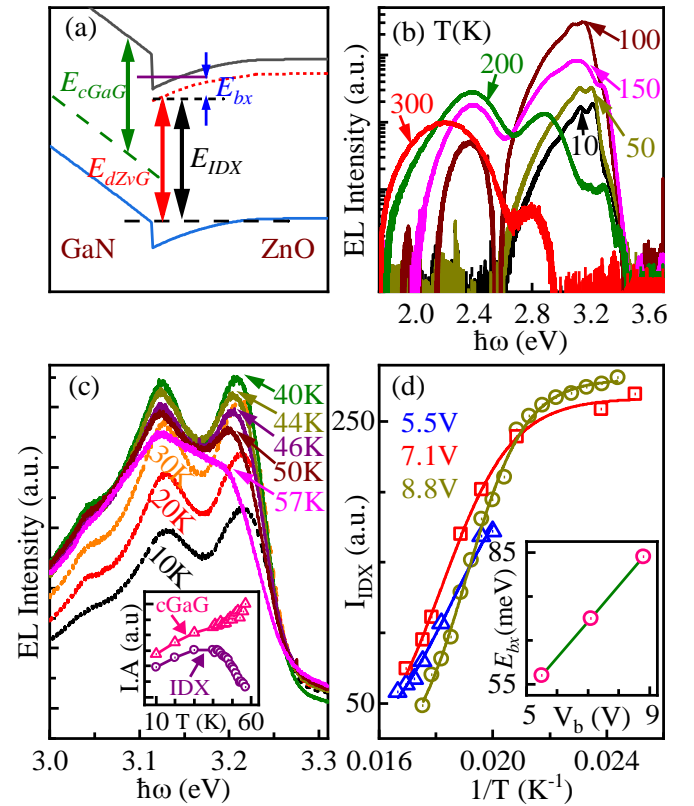

**Fig.4:** (a) Schematic depiction of different transitions taking place at the interface and inside the GaN layer. (b) EL spectra recorded for $V_b = 8.8$V at different temperatures in sample D1. (c) EL spectra recorded at different temperatures below 60K for $V_b = 8.8$V. Inset: integrated intensity of the $E_{cGaG}$ and IDX features versus temperature. (d) IDX intensity versus inverse of temperature for different bias voltages. Inset: binding energy ($E_{bx}$) of the excitons as a function of the applied bias.

Electroluminescence properties of c-oriented n-ZnO/p-GaN heterojunctions are studied as functions of applied bias and temperature. The study evidences the formation of indirect excitons (IDX) at the interface, when the temperature is sufficiently low. The corresponding EL peak shows a systematic redshift with increasing forward bias. Binding energy of these electron-hole coupled entities is found to be quite high. The value can be enhanced to even more than that of ZnO by increasing the applied bias up to a certain limit. Beyond that point, formation of these excitons becomes unsustainable, which can be attributed to the leakage of electrons (holes) into GaN (ZnO) side. Band diagram and the binding energy of these indirect excitons are also theoretically calculated. Theoretical results are found to represent the experimental data very well.

The authors acknowledge financial support from the Department of Science and Technology (DST), Government of India, under Grant No: CRG/2018/001343. They would also like to acknowledge the use of various facilities under the Industrial Research and Consultancy Centre (IRCC), Sophisticated Analytical Instrument Facility (SAIF) and the Centre for Excellence in Nanoelectronics (CEN), IIT Bombay.

**Supplementary Material**

# Spatially indirect interfacial excitons in n-ZnO/p-GaN heterostructures

**Simran, Subhabrata Dhar***

Department of Physics, IIT Bombay, Mumbai-Maharashtra-400076, India.

*E-mail: dhar@phy.iitb.ac.in

# S1. Growth, structural and morphological properties of the ZnO layer

Growth Process: ZnO films were deposited on (0001)-GaN/sapphire substrates using a high vacuum pulsed laser deposition (PLD) system. More details about the ZnO pellet making and growth process can be found elsewhere[1]. A KrF excimer laser with a wavelength of 248 nm and a pulse duration of 30 ns was used to ablate the ZnO pellet target. The laser beam was focused onto the rotating target, which was kept at a distance of 5 cm from the substrate. Pulse energy and frequency of the laser were set at optimized values of 1.5 J cm$^{-2}$ and 5 Hz, respectively. The base pressure of the chamber was below $5 \times 10^{-6}$ mbar. Prior to the deposition, the substrate was cleaned by ultrasonication in trichloroethylene, acetone, and methanol for 5 min each and then immersed in dilute hydrofluoric acid (1:10 ratio) for 1 min before being rinsed in methanol and dried in a $N_2$ flow. Here, we have highlighted two samples as sample D1 and D2 which were deposited at a growth temperature of 500ºC and at oxygen chamber pressures of 15 and 10 mbar on the GaN templates with $n_h \sim 1 \times 10^{17}$ and $1 \times 10^{16}$ cm$^{-3}$, respectively. Growth time was adjusted to 12500 pulses, which corresponds to 41.66 min for both the samples. After the growth, samples were cooled naturally to 100°C. Throughout the cooling stage, oxygen pressure was maintained at the level that was used during growth. The sample was removed from the chamber at room temperature.

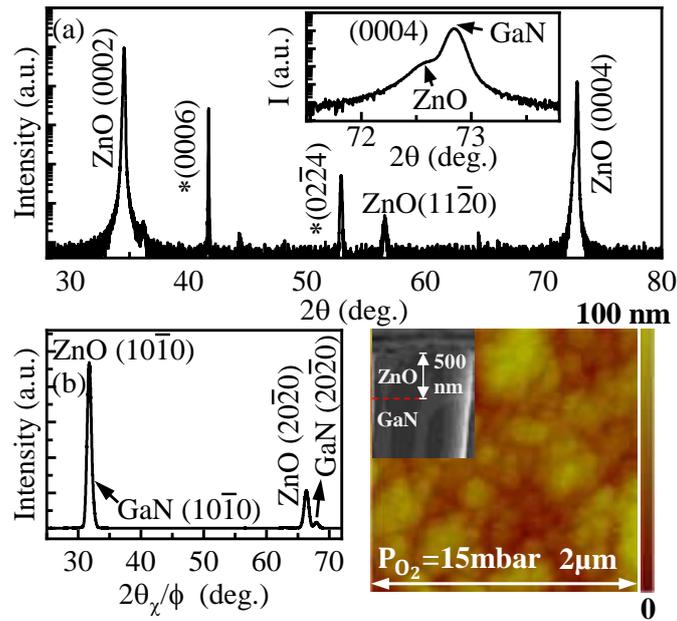

**Fig.S1** High-resolution x-ray diffraction measurements for the samples have been carried out in a Rigaku Smart Lab diffractometer with Cu Kα x-ray radiation. (a) ω-2θ profile recorded for the sample D1, which is featured by (0002) peak and its higher order (0004) reflection for the wurtzite phase of ZnO and GaN (template). It should be noted that these peaks are expected to appear at almost the same position for ZnO and GaN as the lattice constants of the two materials are very similar. The (0006) peak associated with the sapphire substrate is also visible in the scan. Inset shows the (0004) peak in the expanded scale, where the ZnO and GaN features can be clearly distinguished. (b) $2\theta_\chi - \phi$ profile recorded for the same sample. It is featured by (10$\bar{1}$0) and (20$\bar{2}$0) reflections from both ZnO and GaN. The (10$\bar{1}$0) peak for ZnO almost coincides with that of GaN. However, at the (20$\bar{2}$0) reflection, ZnO and GaN features can be clearly distinguished. All these results show the epitaxial growth of (0001) ZnO film on top of (0001) GaN template. $a$ and $c$ lattice parameters are found to be 5.204 and 3.263Å from the XRD peak positions. These values match quite well with those of bulk ZnO[2–4]. This suggests that the grown ZnO layer is unstrained. Note that similar results are obtained for all the samples investigated here. (c) Atomic force micrograph recorded for the ZnO surface of sample D1. A smooth and continuous deposition is quite evident. The rms roughness is found to be ~19 nm for both the samples. Inset: cross-sectional SEM image for sample D1. Thickness of the ZnO film is found to be ~500 nm for both the samples investigated here.

**S2. Current-voltage characteristics**

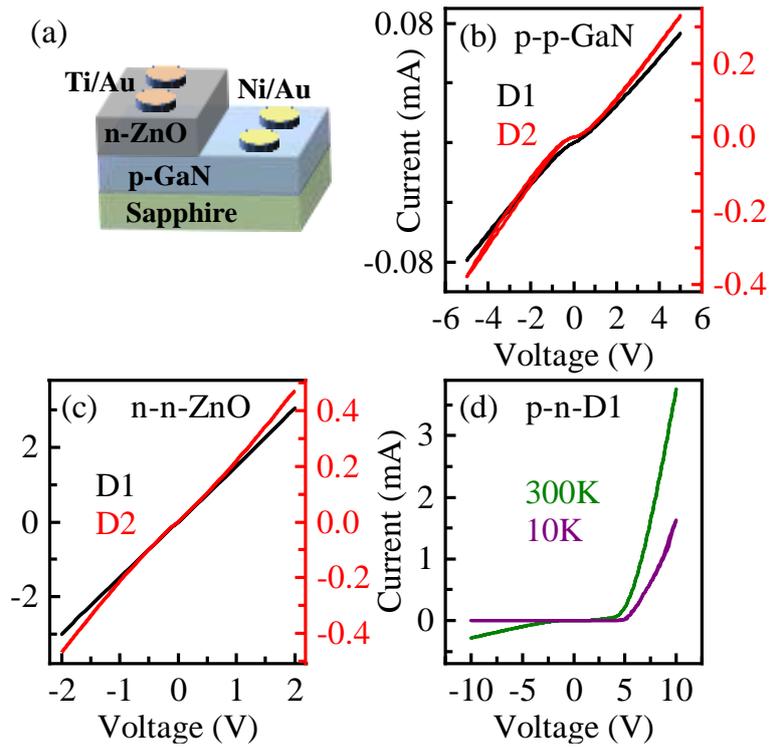

**Fig.S2** (a) Schematic depiction of the n-ZnO/p-GaN heterojunction device. Current-voltage (I-V) profiles recorded at 300K between (b) Ni/Au contacts on p-GaN side and (c) Ti/Au contacts on n-ZnO side for both the devices. While on ZnO, contacts are showing a clear ohmic nature. On GaN side, the profiles are almost linear. (d) I-V profiles recorded between the contacts on GaN and ZnO sides at 10 and 300K for device D1. Both the profiles display rectifying behaviour implying the formation of a depletion region at the interface.

## S3. Comparison of electroluminescence (EL) and photoluminescence (PL) spectra

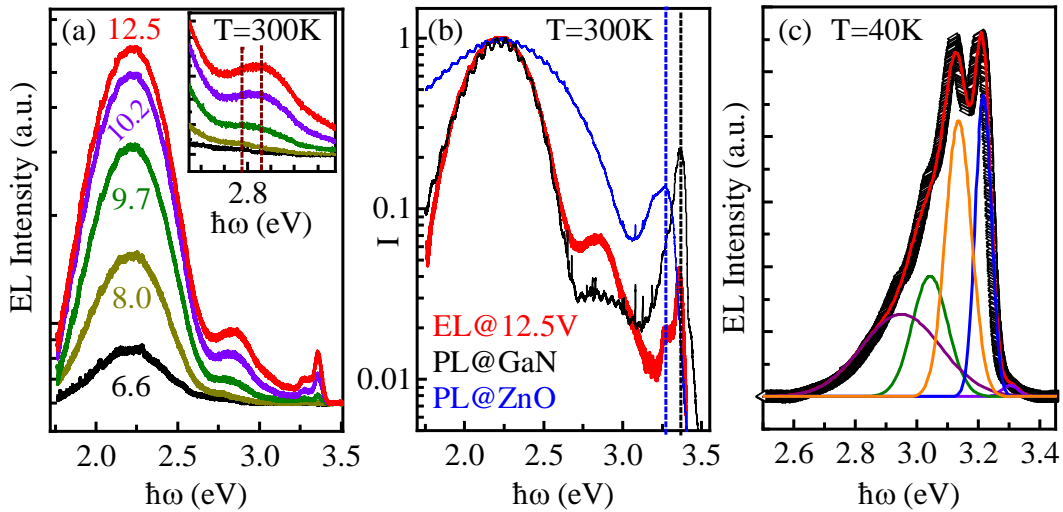

**Fig.S3.1:** (a) EL spectra recorded for the device D1 at 300K at different applied forward voltages. Inset shows the 2.85 eV EL peaks in expanded scales. (b) Compares the EL spectrum obtained at 12.5 V of forward bias with the room-temperature PL spectra recorded separately on ZnO and GaN. EL spectra shown in (a) are featured by four peaks appearing at 2.21 eV, 2.85 eV, 3.27 eV and 3.35 eV. It is quite evident from (a) and (b) that the peaks observed at 3.27 eV and 3.35 eV in EL are resulting from near band-edge emissions in ZnO and GaN, respectively. While the most dominant 2.21 eV EL feature matches very well with the defect related yellow luminescence (YL) feature in GaN PL and hence can be attributed to YL transition in GaN. 2.85 eV EL feature, which appears at a slightly red shifted position as compared to the blue luminescence (BL) PL peak at 2.9 eV in GaN. Note that the 2.85 eV EL peak shows a blue shift with the increase of the forward bias as shown in the inset of (a). These findings may suggest that the origin of the 2.85 eV EL peak is the donor-acceptor pair (DAP) recombination from ZnO donor to GaN acceptor states at the junction (c) EL spectra recorded for 8.8V applied forward bias at 40K temperature for same device. EL feature can be deconvoluted with five gaussian functions. Integrated intensity and peak positions for the two peaks (~3.21 eV and ~3.12 eV) could be obtained from this fitting.

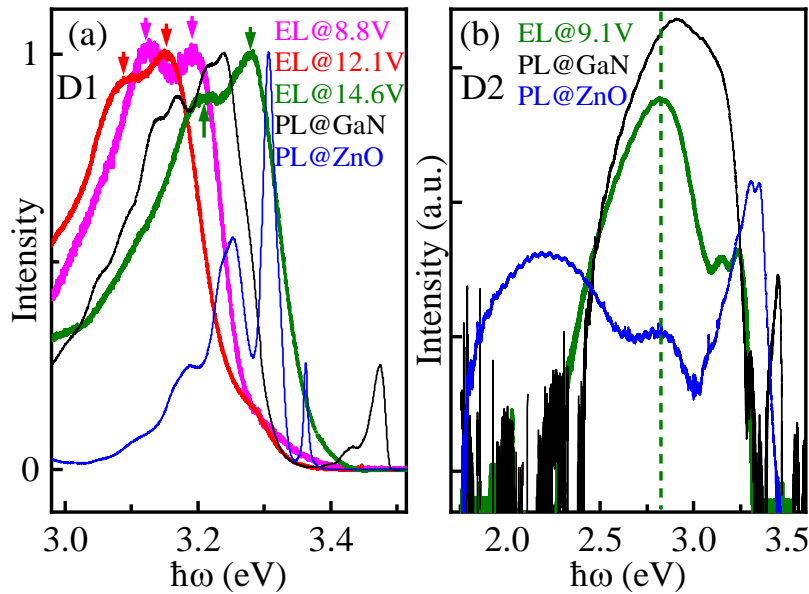

**Fig.S3.2:** (a) PL spectra of ZnO layer and GaN substrate recorded at 10K are compared with the 10K EL spectra for sample D1 at bias voltages belonging to three regimes; 8.8V (regime- I), 12.1V (regime-II) and 14.6V (regime-III). Note that none of the EL peaks appearing above 3 eV band matches with any of the near band edge features of either GaN or ZnO in any of the regimes, which strongly suggests that these EL transitions must be originating from the interfacial region. (b) 10K EL profile for sample D2 recorded under $V_b = 9V$ for the device and PL spectra recorded separately on the ZnO film and GaN substrate at 10K. The most dominant EL feature appearing at 2.78 eV in this sample can be assigned to broad blue luminescence (BL) band observed in GaN PL (also often reported in low temperature PL of Mg doped GaN). However, one must note the line-shape of the EL feature is seemingly quite different from that of the PL. Specially, the 3.23 and 3.14 eV EL features cannot be seen in GaN PL. This may imply that these EL features must also be originating from the interface like in case of device D1.

## S4. Calculation of the band diagram:

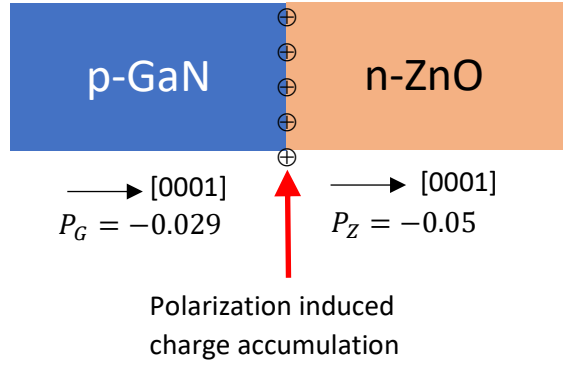

**Fig.S4:** Schematic representation of polarization charge accumulation at the interface of (0001) ZnO layer grown on (0001) GaN.

Band diagrams can be obtained by solving Poisson equation

$$\frac{d^2V}{dx^2} = -\frac{\rho}{\epsilon} \quad \Rightarrow \quad \frac{d^2E(x)}{dx^2} = \frac{q\rho}{\epsilon} \quad (S4.1)$$

where $q$ the electronic charge, $\epsilon$ the dielectric permittivity of the material, $E(x)$ the conduction (valence) band minimum (maximum) as a function of position $x$.

In GaN side, $\rho = q[p(x) - N_{aG}^-]$, where $p(x) = N_{vG}F_{1/2}(\eta)$ the hole concentration as a function of position, $N_{vG} = 2(m_{pG}k_BT/2\pi\hbar^2)^{3/2}$ the effective density of state of the valence band, $m_{pG}$ the hole effective mass, $\eta = (E_{vG}(x) - E_{FG})/k_BT$, $E_{vG}(x)$ and $E_{FG}$ stand for valence band maximum and fermi level in GaN side, $N_{aG}^- = N_{aG}/(1 + g_{va}\exp(\eta)\exp(\Delta E_{aG}/k_BT))$ the ionized acceptor concentration and $\Delta E_{aG}$ the activation energy of the shallow acceptors and $g_{va} = 2$. In the neutral region, electric field should be zero and hence $(dE_{vG}/dx)_{(neutral)} = 0$. Conduction band minimum in GaN side as a function of position can be expressed as $E_{cG}(x) = E_{vG}(x) + E_{gG}$, where $E_{gG}$ the band gap of GaN.

In ZnO side, at the junction, $E_{cZ}(j) = E_{cG}(j) - \Delta E_c$, where $j$ stands for the junction point, $E_{cZ}$ and $\Delta E_c$ represent conduction band minimum in ZnO and conduction band off-set at the interface, respectively. $\rho = q[n(x) - N_{dZ}^+]$, where $n(x) = N_{cZ}F_{1/2}(\eta)$ the electron concentration as a function of position, $N_{cG} = 2(m_{nz}k_BT/2\pi\hbar^2)^{3/2}$ the effective density of state of the conduction band, $m_{nZ}$ the electron effective mass in ZnO, $\eta = (E_{FZ}(x) - E_{cZ})/k_BT$, $E_{cZ}(x)$ and $E_{FZ}$ stand for conduction band minimum and fermi level in ZnO side, $N_{dZ}^+ = N_{dZ}/(1 + g_{cd}\exp(\eta)\exp(\Delta E_{dZ}/k_BT))$ the ionized donor concentration and $\Delta E_{dZ}$ the activation energy of the shallow donors. $g_{cd} = 2$. Furthermore, the condition $D_{jZ} = (P_G - P_Z) + D_{jG}$ should be satisfied at the interface, where $D_{jG}$ and $D_{jZ}$ represent the displacement field at the junction in the GaN and ZnO sides, respectively. $P_G$ and $P_Z$ stand for the spontaneous polarization, respectively, in GaN and ZnO. In the neutral region, electric field should be zero and hence $(dE_{cZ}/dx)_{(neutral)} = 0$. Valence band maximum in the ZnO side as a function of position can be expressed as $E_{vZ}(x) = E_{cZ}(x) - E_{gZ}$, where $E_{gZ}$ the band gap of ZnO. Note that under an applied voltage of $V_j$ the condition $qV_j = E_{FZ} - E_{FG}$ is satisfied.

Equation (S4.1) is solved numerically satisfying above mentioned boundary/interfacial conditions. Parameters used in these calculations are listed in Table S4.1. Note that the temperature dependent variation of band gap in the two materials has been considered as $E_g(T) = E_g(0) - \gamma T^2/(T + \beta)$.

**Table S4.1:** Material parameters used for the calculation

|  | GaN | ZnO |
|---|---|---|
| Band gap at 0K ($E_g$) | 3.47 eV [A] | 3.42 eV [E] |
| Temp. coefficients of $E_g$ ($\gamma, \beta$) | 7.2× $10^{-4}$ eVK$^{-1}$, 600K [B] | 5.3× $10^{-4}$ eVK$^{-1}$, 330K [F] |
| Hole effective mass ($m_p$) | 1.4 $m_o$ [A] | 0.59 $m_o$ [E] |
| Electron effective mass ($m_n$) | 0.2 $m_o$ [A] | 0.24 $m_o$ [E] |
| Dielectric constant ($\kappa$) | 9.5 [C] | 8.5 [E] |
| Spontaneous polarization ($P$) | −0.029 C/m² [D] | −0.05 C/m² [G] |
| Donor concentration ($N_d$) | $1 \times 10^{13}$ cm$^{-3}$ | $1 \times 10^{19}$ cm$^{-3}$ |
| Acceptor concentration ($N_a$) | $1 \times 10^{17}$ cm$^{-3}$ and $2 \times 10^{18}$ cm$^{-3}$ | $1 \times 10^{13}$ cm$^{-3}$ |
| Donor activation energy ($\Delta E_d$) | 15 meV | 30 meV |
| Acceptor activation energy ($\Delta E_a$) | 140 meV [A] | 180 meV |
| Conduction band off-set ($\Delta E_c$) | 200 meV | |

[A] Bougrov V. et al., in Properties of Advanced Semiconductor Materials GaN, AlN, InN, BN, SiC, SiGe., John Wiley & Sons, Inc., New York, 2001, 1-30.

[B] M. Ilegems et al., Journal of Applied Physics **43**, 3797 (1972).

[C] Barker et al., Infrared Lattice Vibrations and Free-Electron Dispersion in GaN, Phys. Rev. B **7,** 743 (1973).

[D] J. Lähnemann et al., Phys. Rev. B **86**, 081302 (2012).

[E] D.P. Norton et al., Mater. Today **7**, 34 (2004).

[F] R.C. Rai et al., Journal of Applied Physics **111**, 073511 (2012).

[G] A. Dal Corso et al., Phys. Rev. B **50**, 10715 (1994).

## S5. Calculation of binding energy of the indirect excitons (IDX):

Binding energy $E_{bx}$ of such excitons can be obtained by solving the Schrodinger equation in cylindrical coordinate[5]. The radial part of the wavefunction $R(\rho)$ satisfies

$$\rho^2 \frac{d^2R}{d\rho^2} + \rho \frac{dR}{d\rho} + \left[\frac{2\mu\rho^2}{\hbar^2}\{E_{nl} - V(r)\} - l^2\right] R(\rho) = 0 \qquad (S5.1)$$

where, $\rho$ represents the cylindrical coordinate for the relative position of the electron and hole. $E_{nl}$ the eigen energy for the principle quantum number $n$ and angular momentum quantum number $l$. $\mu = m_e + m_h/m_e m_h$ the reduced mass of the exciton. Electron hole interaction potential can be considered as $V(r) = -\alpha e^2/4\pi\epsilon_o \kappa_{av}\sqrt{\rho^2 + d_x^2}$, where $\kappa_{av} = (1/2)(\kappa_{GaN} + \kappa_{ZnO})$ the average of the dielectric constants of GaN and ZnO, $\epsilon_o$ the vacuum permittivity, $d_x$ is the average electron-hole separation along the vertical direction and $\alpha$ is a constant introduced to take into account the overestimation of dielectric screening effect in GaN and ZnO. While the azimuthal part of the wavefunction satisfies $\Phi(\varphi) = e^{-il\varphi}/\sqrt{2\pi}$, where $l = -n + 1, -n + 2, \dots, 0, \dots, n - 2, n - 1$ for the principle quantum number $n$. Equation (1) can be solved using Mathematica's 'NDEigensystem' function to obtain the binding energy of the excitons $E_{bx}$. Note that $E_{bx} = E_{nl}$, when $n = 0$ and $l = 0$.